\documentclass[sigconf,nonacm]{acmart}

\begin{document}

\title[Playsemble: Learning Low-Level Programming Through Interactive Games]{Playsemble: Learning Low-Level Programming Through Interactive Games}

\author{Elliott Wen}
\orcid{0000-0002-0340-9392}
\email{elliott.wen@auckland.ac.nz}
\affiliation{%
  \institution{The University of Auckland}
  \city{Auckland}
  \country{New Zealand}
}

\author{Paul Denny}
\orcid{0000-0002-5150-9806}
\email{paul@cs.auckland.ac.nz}
\affiliation{%
  \institution{The University of Auckland}
  \city{Auckland}
  \country{New Zealand}
}

\author{Andrew Luxton-Reilly}
\orcid{0000-0001-8269-2909}
\email{a.luxton-reilly@auckland.ac.nz}
\affiliation{%
  \institution{The University of Auckland}
  \city{Auckland}
  \country{New Zealand}
}

\author{Sean Ma}
\orcid{0000-0002-3350-004X}
\email{sean.ma@auckland.ac.nz}
\affiliation{%
  \institution{The University of Auckland}
  \city{Auckland}
  \country{New Zealand}
}

\author{Jun Seo}
\orcid{0000-0001-5899-2282}
\email{jun.seo@auckland.ac.nz}
\affiliation{%
  \institution{The University of Auckland}
  \city{Auckland}
  \country{New Zealand}
}

\author{Bruce Sham}
\orcid{0000-0001-7007-6746}
\email{bruce.sham@auckland.ac.nz}
\affiliation{%
  \institution{The University of Auckland}
  \city{Auckland}
  \country{New Zealand}
}

\author{Yu Yang}
\orcid{0000-0001-9354-3909}
\email{yangyy@eduhk.hk}
\affiliation{%
  \institution{The Education University of Hong Kong}
  \city{Hong Kong}
  \country{Hong Kong}
}

\author{Chenye Ni}
\orcid{0009-0003-3461-2056}
\email{cni586@aucklanduni.ac.nz}
\affiliation{%
  \institution{The University of Auckland}
  \city{Auckland}
  \country{New Zealand}
}

\begin{abstract}
Teaching assembly programming is a fundamental component of undergraduate computer science education, yet many students struggle with its abstract and low-level concepts. Existing learning tools, such as simulators and visualisers, support understanding by exposing machine states. However, they often limit students to passive observation and provide few opportunities for meaningful interaction. To address these limitations, we introduce Playsemble, a gamified learning system that transforms assembly instructions into interactive, game-like tasks in which students control Pac-Man to collect items, avoid ghosts, and reach targets.
Playsemble integrates a code editor, a multi-architecture CPU emulator, and visual debugging tools within a browser-based environment, allowing students to work offline without installation or configuration. It also provides immediate formative feedback enhanced by large language models. We deployed Playsemble in an undergraduate computer architecture course with 107 students. The course featured a sequence of assignments of increasing complexity, covering core concepts such as register and memory manipulation, control structures including loops and conditionals, and arithmetic operations.
Our findings suggest that Playsemble promotes active experimentation, sustained engagement, and deeper conceptual understanding through meaningful game-based learning experiences.

\end{abstract}


\copyrightyear{2027}
\acmYear{2027} 
\setcopyright{cc}
\setcctype{by}
\acmConference[SIGCSE TS 2027]{Proceedings of the 58th ACM Technical Symposium on Computer Science Education V.1}{February 17--20, 2027}{Sacramento, CA, USA}
\acmBooktitle{Proceedings of the 58th ACM Technical Symposium on Computer Science Education V.1 (SIGCSE TS 2027), February 17--20, 2027, Sacramento, CA, USA}
\acmDOI{10.1145/3857222.3860159}
\acmISBN{979-8-4007-3075-7/2027/02}

\maketitle

\begin{figure*}[h]
\centering
\includegraphics[width=0.95\linewidth]{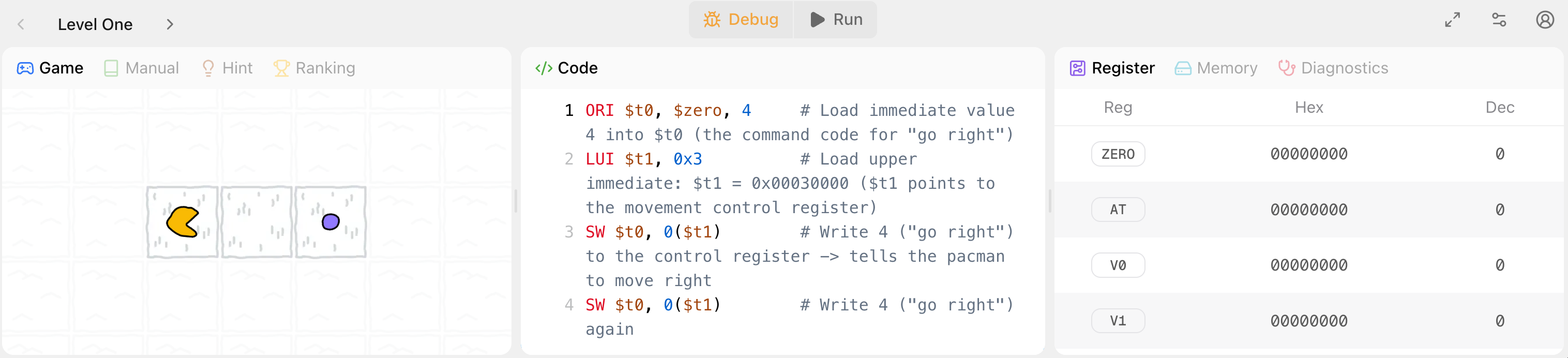}
  \caption{The user interface of Playsemble.}
  \label{fig:screenshot}
\end{figure*}

\section{Introduction}

Teaching assembly programming has long been a cornerstone of the undergraduate computer science curriculum. The ACM/IEEE Computer Science Curricula 2023 report identifies ``Assembly-level Machine Organization'' as a core knowledge unit within the Architecture and Organization area, and uses ``writing a simple assembly language program'' as an illustrative learning outcome~\cite{CS2023}.
Mastery of assembly also remains vital across a range of domains, including robotics and embedded systems~\cite{bolanakis2011teaching}, security analysis~\cite{kongmunvattana2016cybersecurity}, systems programming~\cite{cadenas2015virtualization}, and even artificial intelligence~\cite{tanner2018tensile}.
Yet, students frequently struggle to learn assembly. Its close proximity to the hardware and reliance on low-level instructions for memory and registers often make it feel abstract and difficult to grasp~\cite{worister2025block}.

To address these challenges, prior research has proposed a variety of instructional approaches, with visualisation being the most commonly used. For example, RALPHO~\cite{estep2005flexible} and VisAA~\cite{ly2008visaa} generate control flow diagrams to help students understand program behavior. GSPIM~\cite{borunda2006gspim}, MARS~\cite{vollmar2006mars}, and AsmVisualizer~\cite{newhall2025asm} provide a visual interface that illustrates the dynamic state of a simulated machine, such as registers and memory.
Despite these advances, these tools typically offer limited interactivity; students often passively observe abstract machine states, which are disconnected from meaningful, real-world tasks. They provide little motivation or playful engagement, making learning feel dry and difficult to sustain.

In this study, we introduce Playsemble, a gamified system designed to support the learning of assembly language. Playsemble transforms abstract machine operations into interactive, game-like tasks in which students use assembly instructions to control Pac-Man to collect dots, avoid ghosts, and reach designated targets. A ranking system further motivates learners by rewarding both correctness and efficiency. This design offers meaningful challenges that foster active engagement, sustained motivation, and a deeper understanding of program behaviour. To the best of our knowledge, this is the first study to investigate gamification in the context of assembly language education.

Playsemble integrates a complete set of development and debugging tools within a browser-based environment, such as a code editor, a multi-architecture CPU emulator, and a debugger. The platform is accessible from any device and supports offline use, requiring no installation or configuration. Its debugging interface visualises registers and memory and supports step-wise execution, enabling students to reason about program state and control flow. In addition, Playsemble automatically evaluates submissions and delivers immediate formative feedback.

In this paper, we share our experiences and insights from the first integration of Playsemble into a computer architecture course with 107 undergraduate students. We designed a sequence of programming assignments that progressively connect core assembly concepts to game-based tasks of increasing complexity. Early assignments focus on fundamental skills such as register manipulation and memory access by controlling Pac-Man’s movement in simple environments. Subsequent tasks introduce loops and conditional branching to enable efficient maze traversal and adaptive decision-making. Later stages require students to respond to dynamic elements, such as avoiding moving ghosts, while encapsulating repeated behaviours into functions. These tasks reinforce key concepts, such as parameter passing, stack management, and control flow.
Our findings suggest that Playsemble fosters active experimentation, sustained engagement, and deeper conceptual understanding by grounding abstract assembly concepts in meaningful, interactive game-based tasks.
\vspace{-10pt}
\section{Background}
Previous studies have explored various tools  to support students in learning assembly programming. For instance, Cadenas et al.~\cite{cadenas2015virtualization} introduced a virtual ARM-based system for teaching assembly language. The system integrates essential development tools, such as GNU assemblers and debuggers. Although it provides extensive execution and debugging information, novices may still find the output difficult to interpret and apply effectively. 

To mitigate this challenge, visualisation tools have been developed to present program behaviour in a more intuitive and accessible manner. Tools such as RALPHO~\cite{estep2005flexible}, VisAA~\cite{ly2008visaa}, and PathViz~\cite{ebersole2007visualizing} generate control flow diagrams to help students understand potential jumps, function calls, and program structure. GSPIM~\cite{borunda2006gspim}, MARS~\cite{vollmar2006mars}, Apoo~\cite{reis2001apoo}, and CALVIS32~\cite{alcalde2016calvis32} provide richer details by displaying the dynamic state of the machine, such as register and memory contents, step by step. ASM Visualizer~\cite{newhall2025asm},  MieruCompiler~\cite{gondow2010mierucompiler}, and
Frances~\cite{sondag2012frances} advanced this line of work by offering a browser-based interface that supports multiple assembly languages. 

Building on this line of research, block-based approaches leverage extensive research on the benefits of block-based programming for high-level languages~\cite{xu2019block} by representing assembly instructions as graphical blocks. For instance, Blocksambler~\cite{worister2023block} proposes a block-based assembly-like language, while BRISC-V~\cite{agrawal2019brisc} visualises RISC-V instructions through graphical blocks. These tools reduce syntactic overhead and make program structure more immediately apparent. They also support a smoother transition from conceptual understanding to low-level programming, by enabling learners to manipulate visual blocks instead of raw code.

Both visualisation and block-based tools can enhance comprehension by linking abstract instructions with concrete machine behaviour. However, they largely provide passive learning experiences and offer limited opportunities for meaningful interaction. To address these limitations, our work investigates gamified, interactive systems that integrate learning objectives into engaging, hands-on tasks.  Interactive visual environments have long been used to make programming concepts more tangible. Games such as Lightbot~\cite{gouws2013lightbot} encourage learners to experiment with sequencing and control through immediate visual feedback, while the use of various microworlds that embed code within puzzle-like contexts have been shown to support conceptual exploration and motivation \cite{pelanek2022design}. Similar approaches combining visualisation and gameplay, such as DeCode for data structures \cite{su2021game}, have demonstrated positive effects on student enjoyment and learning.  Building on these ideas, Playsemble extends interactive, game-based learning to the domain of assembly programming.

\section{Design}
Figure~\ref{fig:screenshot} presents the Playsemble user interface, which is organised into three primary columns. The central column contains an editor for writing assembly code to control Pac-Man. The left column displays the Pac-Man game, allowing students to observe how their programs affect its movements in real time. The right column provides a visual debugging interface that dynamically displays memory and register values. It supports stepwise execution and state inspection in both forward and backward directions. The following sections provide a detailed description of the system’s core components.

\noindent{\textbf{Editor:}} The assembly source code editor is built on the Monaco editor from {VS Code}\footnote{https://microsoft.github.io/monaco-editor/} and offers productivity features such as syntax highlighting, autocomplete, and on-hover instruction explanations. A dedicated language server, {asm-lsp}\footnote{https://github.com/bergercookie/asm-lsp}, is integrated to provide additional functionality, such as reference viewing, signature help, and real-time error diagnostics. The editor preserves progress across sessions, ensuring that code remains intact after a browser refresh or system restart. It is also optimised for portable devices with touch-friendly controls, gesture support, and a responsive layout. 

\noindent{\textbf{Assembler:}} When the user clicks the ``Run'' button, the text-based assembly code is assembled into binary instructions. This is achieved using the GNU assembler from GNU Binutils~\cite{pesch1993gnu}, which supports multiple popular assembly languages, such as ARM, x86, and MIPS. 
Once the binary instructions are generated, we further leverage the nm tool\footnote{https://man7.org/linux/man-pages/man1/nm.1.html} to perform a series of compile-time semantic checks, for example, verifying that all symbols are correctly defined and that they are located within the executable memory region. This helps assembly learners identify typical mistakes and correct them early.
We compile both the GNU assembler and nm tools into WebAssembly. This enables all assembly processing to occur locally within the browser.

\noindent\textbf{Virtual CPUs:}
Playsemble provides virtual CPUs to execute the generated binary instructions. These virtual CPUs are implemented using the lightweight CPU emulator framework Unicorn~\cite{quynh2015unicorn}. It provides precise emulation of CPU instruction sets such as x86, ARM, MIPS, and RISC-V. Playsemble compiles Unicorn to WebAssembly, enabling local execution in the browser. This process poses a technical challenge, as Unicorn relies on features that are not fully supported in WebAssembly, for example, variadic function arguments and dynamic code generation. Playsemble addresses these challenges by applying specific modifications, such as generating function wrappers and implementing an inline interpreter, to ensure that Unicorn functions correctly within the browser environment. 
\begin{figure}[t]

\centering
\includegraphics[width=0.75\linewidth]{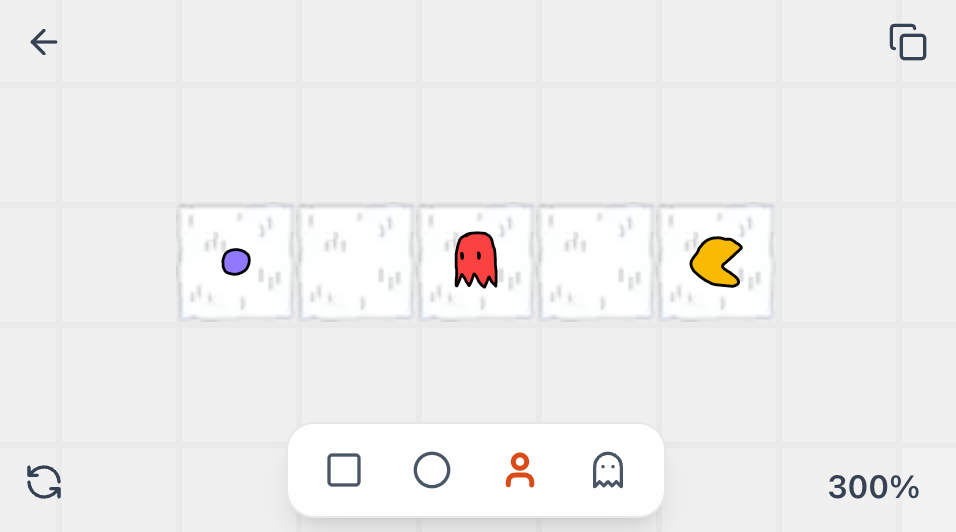}

  \caption{The map builder's user interface enables educators to place, arrange, and manage tiles.}
  \label{fig:map_builder}
  \vspace{-15pt}
\end{figure}

\noindent\textbf{Pac-Man Game:}
We implement a simplified version of the classic Pac-Man game as a platform for assembly programming exercises. 
The game is integrated with our virtual CPUs. It allows the Pac-Man character's movement to be controlled directly by the user's assembly program through memory-mapped registers (i.e., mmio). For example, the memory at \texttt{0x30000} is used to issue movement commands: writing \texttt{1} to the address moves Pac-Man up, while writing \texttt{2} moves it down. A dedicated memory region starting at \texttt{0x30010} allows the user program to access the map data. The map is represented as a byte matrix with $X$ rows and $Y$ columns, where each byte corresponds to a map tile. For example, a tile value of \texttt{0} represents a wall, \texttt{1} represents Pac-Man, and other values may correspond to ghosts or collectible items. The dimensions $X$ and $Y$ can also be read from dedicated memory registers.
We also provide a map builder to enable educators to create custom maps easily, 
as illustrated in Figure~\ref{fig:map_builder}. 
The map builder supports an event handler for individual map tiles. For example, a ghost can be assigned a function implementing an A* search algorithm, making it actively chase Pac-Man. Similarly, collectibles can be configured to appear randomly when the game starts, which adds dynamic elements to the gameplay.

\noindent\textbf{Debug Panel:}  
This component provides real-time visualisation of both registers and memory, allowing students to see the immediate effect of each instruction on the CPU state. To support deeper inspection, students can set breakpoints at specific instructions and pause execution to examine the program in detail. 
The system also offers step-wise debugging in both forward and backward directions, similar to PythonTutor's approach~\cite{hassan2024evaluating}. This bidirectional execution allows learners to replay and revisit program behaviour. Prior research~\cite{xie2023developing} shows that such replay and revisit capability can promote self-regulation by encouraging students to reflect on their problem-solving process and identify areas for improvement.

\section{Experience}
We deployed Playsemble in a computer systems architecture course with 107 third-year undergraduate students. In this section, we report our experiences and lessons learned. The collection and analysis of data were conducted under institutional ethics approval for the research use of student-generated data.

\subsection{Assignment Design}
\begin{figure*}[h]
\centering
\includegraphics[width=0.85\linewidth]{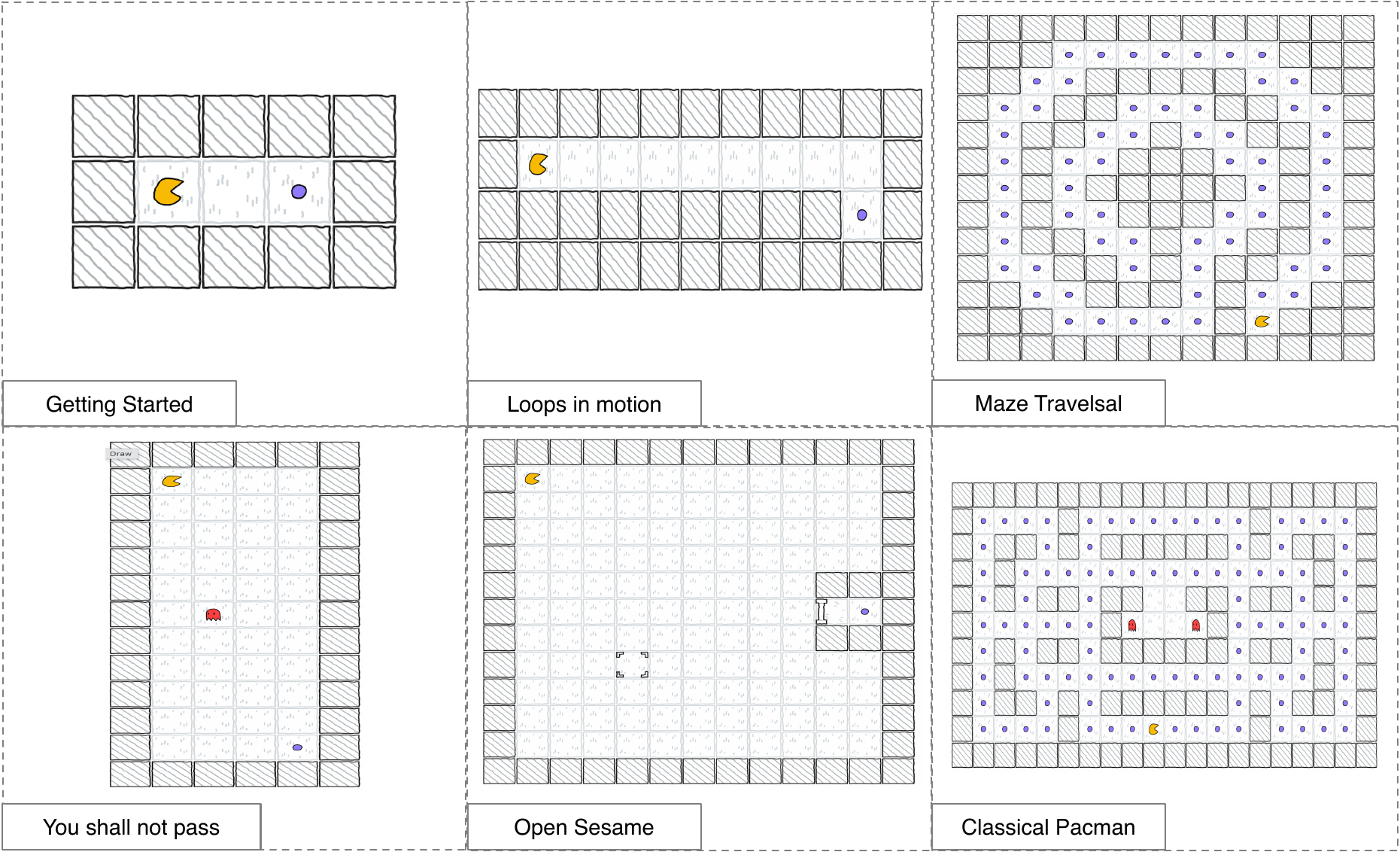}

  \caption{Our assignment is designed to cover the core concepts of assembly programming.}
  \label{fig:stage}

\end{figure*}

The course assessment consists of two equally weighted components: 50\% is based on a theoretical exam, and the remaining 50\% is distributed across five programming assignments, as illustrated in Figure~\ref{fig:stage}. These assignments are carefully designed to cover the core concepts of assembly programming as follows.

\noindent\textbf{Stage 1 --- Getting Started:} The first assignment asks students to control Pac-Man to move right and collect dots located two steps away. It introduces basic register manipulation and memory access. This task can be completed in as few as four lines of code, providing a gentle introduction to the platform. It allows students to build confidence and become familiar with the environment before progressing to more complex tasks.

\noindent\textbf{Stage 2 --- Loops in Motion:} The second assignment requires students to control Pac-Man to collect dots located eight steps ahead and one row below. Students are initially provided with a solution where each movement is written out explicitly. As a result, the same instructions appear many times in the code. Students are then encouraged to refactor this solution using loops. By comparing the two approaches, students can examine how iteration is implemented at the instruction level and how it influences code structure and efficiency.

\noindent\textbf{Stage 3 --- Maze Traversal:}
In the third assignment, Pac-Man starts at the outer edge of a spiral maze and must collect dots while following the path inward toward the center. At each step, students evaluate the map data to determine the correct direction to the next dot. This stage introduces conditional instructions and branching. It moves students from fixed, repetitive sequences to data-driven control flow. It helps students develop the ability to adapt program logic to structured input and to plan sequences of actions based on predictable patterns in the environment.

\noindent\textbf{Stage 4 --- You Shall Not Pass:}
In this stage, Pac-Man must navigate from the first row to collect the dot on the last row. A ghost patrols a middle row, moving randomly left and right. Pac-Man must traverse this row without being captured. This stage focuses on reactive decision-making in an unpredictable environment. Students must plan their movements strategically and adapt in real-time to dynamic obstacles. Compared with stage 3, this stage requires students to apply conditional branching and loops to handle unpredictable, real-time situations. 

\noindent\textbf{Stage 5 --- Open Sesame:}
Pac-Man is trapped behind a locked gate. To unlock it, he must step onto a single Glyph Tile, which appears at a random location in the maze. Upon reaching the tile, he must perform a fixed movement pattern (e.g., move up three steps, right two, then down three). After completing the sequence, the current Glyph Tile disappears and a new one appears elsewhere. Pac-Man must repeat this process a random number of times. Once all sequences are completed, the gate opens, and he can collect the dots hidden behind it.  Because the same movement pattern must be executed multiple times from different starting points, students are naturally led to encapsulate the repeated instructions in a function and invoke it whenever a new Glyph Tile appears. The purpose of this task is to give students a concrete setting in which to examine how functions are implemented in assembly. It introduces key mechanisms such as parameter passing, stack usage, and return control flow, helping students connect familiar high-level abstractions to their assembly-language equivalents.

\noindent\textbf{Optional Stage --- Classic Pac-Man:} In this stage, Pac-Man must collect as many dots as possible on a simplified version of the original map while avoiding two ghosts. The ghosts mimic the behaviour of classical Pac-Man, moving efficiently by selecting, at each intersection, the next tile that minimises the Manhattan distance to the target tile (in our implementation, the target is always Pac-Man). This assignment challenges students to integrate all previously learned techniques to navigate the maze efficiently and safely. It reinforces their ability to manage complex, dynamic tasks in assembly programming. This stage carries no marks and is entirely optional.

\begin{figure}[t]
\centering
\includegraphics[width=0.85\linewidth]{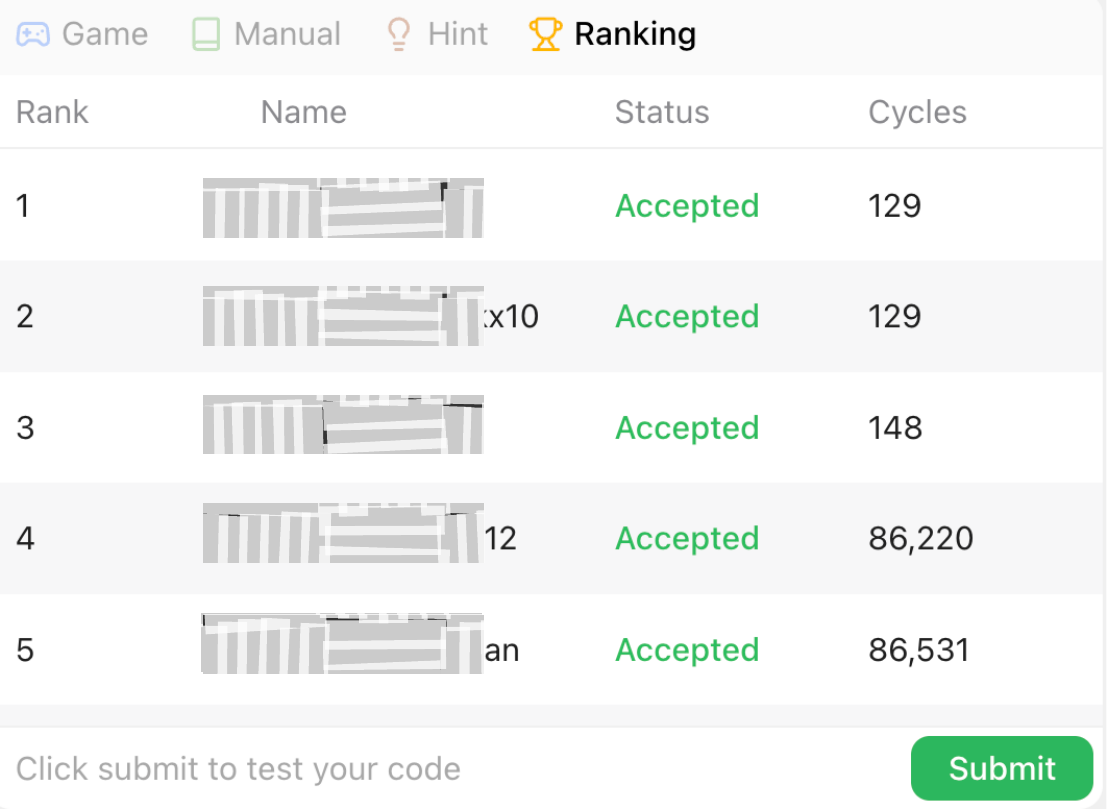}

  \caption{We automatically assess and rank students' submissions based on the CPU cycles.}
  \label{fig:ranking}

\end{figure}

\subsection{Assignment Assessment}
Students are allowed to submit their code to an automated assessment tool to obtain constructive feedback prior to the assignment deadline. The marking system performs two essential checks. First, it verifies whether the assembly code can be successfully assembled into binary instructions. Second, the system executes the assembled program within our virtual CPU to determine whether it satisfies the finishing criteria for the assignment (e.g., collecting all required dots). If the program fails to reach the expected outcome, the assessment tool reports a specific failure reason, such as Pac-Man stopping prematurely or issuing no movement commands.
If both checks pass, the submission is marked as accepted. The system also records the total number of CPU cycles consumed by the program and uses this metric to generate a leaderboard of efficient solutions, as shown in Figure~\ref{fig:ranking}.

When errors occur, the system uses LLM support to generate formative feedback. This aligns with recent work demonstrating that LLMs can produce learner-friendly error messages and code explanations that improve students' understanding and debugging effectiveness~\cite{leinonen2023using, leinonen2023comparing}. We utilize the OpenRouter API to provide access to state-of-the-art LLMs (e.g., \textit{qwen3.6-35b-a3b}). We adopt the CRISPE prompt framework~\cite{wang2023unleashing}. The constant elements of our prompt specify the \textit{Role} (an expert in computer system architecture and a skilled instructor), the \textit{Context} (students learning assembly programming on a gamified Pac-Man platform, supported by a concise platform manual), and the \textit{Personality} (a clear, direct, and instructional communication style).
The variable elements depend on the stage of assessment.
If an error occurs in the first check, the system forwards the assembler output together with the student’s code to the LLM.
If the error occurs during execution, the assessment tool further extracts a compact set of execution signals, such as the last few executed instructions, the current register snapshot, and a small memory slice around recently accessed addresses.

\begin{figure}[t]
\centering
\includegraphics[width=0.95\linewidth]{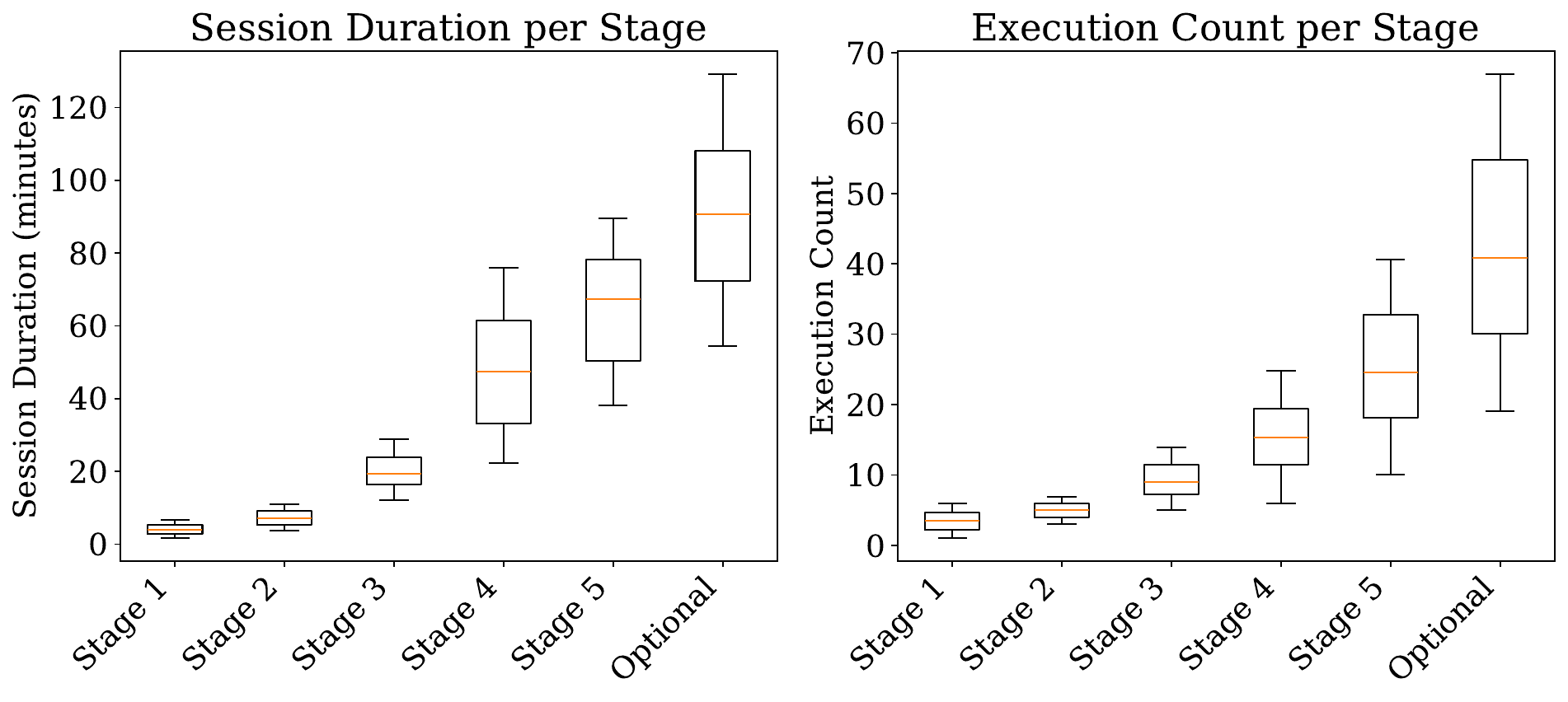}

  \caption{Session Duration and Execution Count for Each Stage}
  \label{fig:usage}

\end{figure}

\subsection{System Usage Analysis}
To understand how students engaged with Playsemble, we analysed system usage data collected throughout the semester. 

\noindent \textbf{Session Duration and Execution Count:} As shown in Fig~\ref{fig:usage}, both session durations and execution counts increased steadily as the stages grew more complex. Early tasks  were typically completed in short sessions with relatively few runs, with median session durations of 3.1 and 8.7 minutes and median execution counts of 3 and 4. These patterns suggest that students were able to quickly experiment with the tasks and familiarise themselves with the platform. In later stages, particularly Stage 4 and Stage 5, students engaged in substantially longer sessions and ran their programs far more frequently, with median durations of 29.8 and 47.1 minutes and median execution counts of 16 and 22. This shift reflects increased engagement in iterative cycles of testing, debugging, and refinement driven by the higher complexity of the tasks. Notably, despite the absence of formal marks, the optional stage also showed extended session durations (median 54.5 minutes) and high execution counts (median 39).

\noindent \textbf{Analysis of Error Patterns:} We analysed students' error logs to identify common difficulties. In the early stages (Stages 1-2), many students were able to submit once and pass the automated checks without requiring further revisions. This suggests that these introductory tasks effectively supported students in quickly grasping fundamental assembly concepts.
As the stages grew more complex, both the frequency and diversity of errors increased. Stage 4 provides a representative example: around 83.7\% of students encountered at least one error in their first submission. Assembly errors arising from syntactic mistakes affected only 16.6\% of users, whereas the majority of errors were logical in nature. 
Several recurring logical error patterns were observed. One common issue was infinite loops. In these cases, students did not update loop variables correctly, causing Pac-Man to stall and stop making progress. Off-by-one and boundary errors were also frequent. Incorrect loop bounds or map indexing often caused Pac-Man to collide with walls or miss target tiles.
Incorrect branch conditions formed another major category of errors. Some students branched on the wrong registers or misunderstood low-level comparison semantics. This caused the program to follow unintended execution paths.

\noindent\textbf{Debugger Usage:} Students made extensive use of the debugging features. Approximately 67.2\% of failed executions were followed by a debugger invocation, and 73.4\% of students used the debugging tools at least once, with usage increasing in the later stages of the assignments. Notably, 21.1\% of students continued to use the debugger even after their programs achieved the correct outcome, particularly when attempting to improve their ranking on the leaderboard.
Students also showed a clear progression in how they used debugging features. In earlier stages, step-wise execution accounted for the majority of debugger interactions, whereas breakpoint usage increased substantially in later stages. Specifically, fewer than a quarter of students employed breakpoints in the early assignments, but this proportion rose to over half in Stages 4. This shift suggests that students developed more targeted and systematic debugging strategies as task complexity increased and their familiarity with the debugging interface grew.

\noindent \textbf{Ranking System:} System logs show that social comparison and friendly competition were strong motivators. In total, 64.7\% of students continued iterating on their code after achieving a successful pass. This suggests that many learners were not satisfied with task completion alone and instead aimed to reduce cycle counts.
Stage 4 provides a representative example. The most active student submitted 342 program versions. Of these, 295 executions occurred after the first successful pass. During this process, the cycle count was reduced from 336,382 to 129. Another student achieved 129 cycles on the first pass but still performed 155 additional runs. This further reduced the execution cost to 118 cycles.

\subsection{Student Feedback}

Student feedback was collected in end-of-course questionnaires that asked students what aspects of the course were `most helpful for your learning?''.

\noindent \textbf{Playful and Intuitive Experience:}  
Several students described Playsemble as making assembly programming feel \textit{``like playing a game rather than writing code''}. One student, who had previously found registers and memory instructions intimidating, shared that \textit{``controlling Pac-Man with assembly made everything click''}. They appreciated how abstract operations were tied to tangible, game-like outcomes. Many students noted that the immediate visual feedback on how their instructions affected Pac-Man's movement allowed them to quickly identify and correct mistakes, saying it let them \textit{``treat every failure as another puzzle to solve''}.

\noindent\textbf{Ranking and Competitive Motivation}  
Many students highlighted the motivational impact of Playsemble's ranking system. Several reported that the leaderboard encouraged deeper engagement with the learning material. As one student noted, \textit{``I kept tweaking my code even after it worked, because I wanted to see how much I could improve my rank.''} In pursuing higher rankings, students experimented with more efficient and creative uses of assembly instructions. For instance, submission logs show that one student replaced conditional branches with arithmetic expressions, encoding the conditional effect directly in a register value and thereby eliminating the need for a branch.

\noindent\textbf{Quality of LLM-generated Feedback:}  
A number of students reported that the feedback generated by the LLM was accurate, timely, and highly useful, particularly when  understanding unexpected behaviours. One student remarked, \textit{``The feedback pointed out exactly where my logic went wrong, and it helped me fix issues that I couldn't see just by stepping through the code.''} For example, one common challenge is determining why a loop or conditional branch does not produce the intended movement for Pac-Man. The LLM was able to pinpoint the issue, such as an incorrect offset in memory access or a misused register, and explain the reasoning behind it. Several students highlighted that the explanations went beyond simple error correction, as the LLM sometimes provides insights into best practices for register usage, instruction sequencing, and memory management. One student noted, \textit{``It didn't just fix the bug for me. It also explained function call conventions and why certain registers need to be saved and restored.''}

\subsection{Lessons Learned by Instructors}

\noindent\textbf{Stage Design Oversight or a Feature?} In Stage 4, a ghost patrols one of the middle rows, moving randomly left and right, and Pac-Man must cross this row without being captured. Our intention was to encourage students to use the loop and condition checking in assembly.   During evaluation, we noticed that some students’ programs consumed significantly fewer CPU cycles than others, which was unexpected. Upon reviewing their code, we discovered a design oversight: the ghost always spawns in the second column and, upon reaching a dead end, consistently turns back. Students could exploit this predictable behaviour by just counting the number of turns and passing safely on even-numbered turns. Interestingly, many students appreciated this quirk, interpreting it as a deliberate hidden shortcut rather than an oversight. As one student commented, \textit{``I felt like I had discovered an Easter egg in the assignment when I figured out the ghost's pattern.''} This sense of discovery transformed an unintended flaw into an engaging and rewarding learning moment.

\noindent\textbf{LLM Limitations.}  
Despite their usefulness, LLMs exhibit some limitations. They may hallucinate, for example by suggesting jumps to non-existent labels. Even when explicitly instructed that the assignments use MIPS Assembly Release~6, the model occasionally generates instructions from other MIPS variants. For instance, it sometimes recommends branch instructions with delay slots, which are not supported in Release~6. These issues highlight the continued importance of instructors in helping students interpret LLM-generated feedback and understand its limitations. It is also important to emphasise that the LLM is an optional enhancement: core learning is driven by scaffolded activities introducing assembly concepts, while the LLM serves only as supplementary support.

\noindent\textbf{Agentic AI and Leaderboard.} An unexpected observation was the extent to which two students leveraged browser-based AI agents to improve their leaderboard rankings in the optional stage. Unlike traditional uses of LLMs for code generation or debugging assistance, these agents were able to directly and autonomously interact with the Playsemble environment, execute programs, analyse outcomes, and iteratively refine solutions. The students reported substantial improvements in their rankings through this process. However, they also noted that the approach incurred a considerable number of tokens and was therefore costly. This observation raises interesting questions about the future role of AI agents in programming education; in particular, educators may need to consider how curricula and assessment designs can adapt to a setting where students routinely collaborate with autonomous agents. At the same time, the reported token costs suggest that human understanding and strategic guidance remain important components of effective agent-assisted problem solving.

\section{Conclusion}
In this paper, we present Playsemble, a gamified platform for learning assembly programming. The platform turns assembly instructions into interactive, game-like tasks. Students use code to control Pac-Man to collect items, avoid obstacles, and reach targets. Playsemble integrates a code editor, a CPU emulator, and a visual debugger within a modern web browser.
The platform also provides immediate formative feedback, supported by a large language model and a ranking system, to encourage active experimentation and engagement. We report on the deployment of Playsemble in an undergraduate course with 107 students. This paper presents a case study based on instructor observations and student feedback. Our findings suggest that game-based activities can effectively support students’ understanding of low-level programming and offer useful guidance for educators designing similar learning environments.

\balance
\bibliographystyle{ACM-Reference-Format}
\bibliography{article-full}

\end{document}